\documentclass[conference]{IEEEtran}
\IEEEoverridecommandlockouts
\usepackage{amsmath,amssymb,amsfonts}
\usepackage{algorithmic}
\usepackage{textcomp}
\usepackage{xcolor}
\usepackage{multirow, makecell}
\usepackage{balance}
\usepackage{graphicx,multicol}
\usepackage[lofdepth,lotdepth]{subfig}
\usepackage{float}
\usepackage{pifont}
\usepackage[backend=biber,style=ieee]{biblatex} 
\newcommand{\para}[1]{\vspace{2mm}\noindent\textbf{#1}}

\newcommand{\matr}[1]{\mathbf{#1}}
\newcolumntype{?}{!{\vrule width 1pt}}
\usepackage[utf8]{inputenc}

\usepackage{hyperref} 
\usepackage{setspace}

\begin{document}


\title{My friends also prefer diverse music: homophily and link prediction with user preferences for mainstream, novelty, and diversity in music
}

\author{\IEEEauthorblockN{Tomislav Duricic\textsuperscript{1,2}, Dominik Kowald\textsuperscript{1}, Markus Schedl\textsuperscript{3,4}, Elisabeth Lex\textsuperscript{2}}
\IEEEauthorblockA{\textit{\textsuperscript{1}Know-Center GmbH, Graz, Austria; \textsuperscript{2}Graz University of Technology, Graz, Austria;} \\ \textit{\textsuperscript{3}Johannes Kepler University Linz, Linz, Austria, and \textsuperscript{4}Linz Institute of Technology, Linz, Austria} \\
tduricic@know-center.at, dkowald@know-center.at, markus.schedl@jku.at, elisabeth.lex@tugraz.at}
}

\maketitle

\begin{abstract}

Homophily describes the phenomenon that similarity breeds connection, i.e., individuals tend to form ties with other people who are similar to themselves in some aspect(s). The similarity in music taste can undoubtedly influence who we make friends with and shape our social circles.
In this paper, we study homophily in an online music platform Last.fm regarding user preferences towards listening to mainstream (M), novel (N), or diverse (D) content. Furthermore, we draw comparisons with homophily based on listening profiles derived from artists users have listened to in the past, i.e., artist profiles. 
Finally, we explore the utility of users' artist profiles as well as features describing M, N, and D for the task of link prediction.
Our study reveals that: (i) users with a friendship connection share similar music taste based on their artist profiles; (ii) on average, a measure of how diverse is the music two users listen to is a stronger predictor of friendship than measures of their preferences towards mainstream or novel content, i.e., homophily is stronger for D than for M and N; (iii) some user groups such as high-novelty-seekers (explorers) exhibit strong homophily, but lower than average artist profile similarity; (iv) using M, N and D achieves comparable results on link prediction accuracy compared with using artist profiles, but the combination of features yields the best accuracy results, and (v) using combined features does not add value if graph-based features such as common neighbors are available, making M, N, and D features primarily useful in a cold-start user recommendation setting for users with few friendship connections.
The insights from this study will inform future work on social context-aware music recommendation, user modeling, and link prediction. 

\end{abstract}

\begin{IEEEkeywords}
Homophily, Link prediction, Computational social science, Modeling user preferences, Online music platform, Cold-start recommendation
\end{IEEEkeywords}

\section{Introduction}


With the emergence of online music streaming platforms, such as Last.fm, Spotify, or Pandora, music has never been more accessible. In addition, many music platforms integrate a social component, allowing users to connect online. 
One of the most critical social mechanisms that drive users to connect is \textit{homophily}, i.e., the tendency of people to connect to others who are similar to themselves in some aspect. Such aspects include sociodemographic, behavioral, or intrapersonal characteristics~\cite{block2014multidimensional}, but also social status, opinions, and values~\cite{lazarsfeld1954friendship}. In the context of music, research has shown that people are attracted to those who share their music preferences~\cite{selfhout2009role}. Besides, similarity in music taste is associated with closeness and relationship satisfaction~\cite{zillmann1989effects}. 

The music taste of individuals is often described by measuring preferences towards specific artists or genres. In addition, users can also be described by their preferences towards mainstream, diverse, or novel music content~\cite{bauer2019global,vigliensoni2016automatic}. Using features capturing those preferences for user modeling has proven effective in a music recommendation setting where traditional music recommendation algorithms tend to provide recommendations of lower quality to users who prefer non-mainstream or novel music~\cite{schedl2015tailoring,kowald2021support}. 

However, it is still unclear to what extent these features impact friendship formation in online music platforms. 

\para{Objective and approach.} 
The study at hand aims to answer the following questions:
(i) Do connected users have similar music tastes for artists they have listened to in the past (i.e., their artist profiles)? We tackle this question by calculating artist profile similarity for connected users in the Last.fm~\cite{schedl2016lfm} dataset with a friendship network and comparing it with users connected in a simulated random network.
(ii) To which extent is friendship formation driven by homophily based on user mainstreaminess (M), novelty (N), and diversity (D)? Here, we compute homophily using the assortativity coefficient~\cite{newman2003mixing}, which is a correlation measure for how similar are users to their neighbors. 
(iii) In case of observable homophily for M, N, D, or artist profile similarity, does it hold over the entire value distribution or are there differences between users depending on their M, N, and D values? 
To answer this question, we categorize users in three different groups for M, N, and D (low/mid/high) based on values of features with highest assortativity coefficients and analyze the ratios between observed and expected edges (O/E ratio assuming random pairing) within and between groups. 
(iv) 
What is the merit of M, N, and D as well as users' artist profiles for friendship prediction? To demonstrate this, we conduct a number of link prediction experiments. We use a binary classification approach with an XGBoost classifier for different combinations of these features and compare it to 
 a weak stratified random baseline and a strong baseline, for which we use an XGBoost classifier with graph-based features. Additionally, we explore feature importance scores of individual as well as aggregated features.

\para{Contributions and findings.} We summarize our contributions as follows. In our experiments, we confirm the existence of homophily based on artist profile similarity and based on the users' preferences for mainstream (M), novel (N), and diverse (D) music. Our experiments show that homophily based on D is the most notable, for example, users who prefer diverse music tend to form friendship connections. Furthermore, we systematically investigate the interplay between the (low/mid/high) values of user mainstreaminess, novelty, diversity, artist profile similarity, and friendship connections between users. We find that some user groups such as high-novelty-seekers exhibit strong homophily but lower than average artist profile similarity. In other words, they tend to form friendship connections even though their listening history is not as similar as with some other user groups. Finally, we investigate the merit of users' preferences for M, N, and D for link prediction and demonstrate that we can outperform the random baseline and achieve similar results as when we employ artist profile features for the same task. The combination of those features results in the best performance. 
To foster reproducibility of our results, we provide our dataset that consists of $11\!,792$ Last.fm users with friendship connections and numeric features describing each user's artist profile, M, N, and D values on Zenodo\footnote{DOI: 10.5281/zenodo.5585638} and the entire Python-based implementation can be accessed via GitHub\footnote{\url{https://github.com/tduricic/homophily-lastfm}}. 

\section{Related Work}\label{sec:related-work}

Several studies exist on homophily and link prediction in online music platforms. A few studies investigate homophily for demographic, topological (graph-based), and taste-related attributes along with friendship connections from Last.fm users and the usefulness of these attributes for the task of link prediction~\cite{bischoff2012we, baym2009tunes}. Findings in those studies show a notable country- and age-based homophily, and a tendency for mixed-gender friendships. However, this tendency is reversed to same-gender friendships if event co-attendance is considered instead of friendship connections. 
When it comes to taste similarity, results in~\cite{bischoff2012we, baym2009tunes} show higher similarity for friends than for random user pairs (also confirmed in ~\cite{aiello2012friendship, zhou2017homophily, guidotti2019know}), but low similarity overall, likely due 
to the sparseness of considered taste profile vectors as discussed in~\cite{bischoff2012we}. Both~\cite{bischoff2012we} and ~\cite{aiello2012friendship} show that graph-based features (e.g., common friends) are most indicative for link prediction. Other important features for link prediction in~\cite{bischoff2012we} are top listened artists, coming from the same country or the number of online interactions between two users. Another study on Last.fm~\cite{bisgin2010does} runs community detection algorithms and compares the averaged user interests of extracted communities to the entire population. Their findings show that on average there are no notable differences in user interests between extracted communities, i.e., users in a detected community have no distinctive preference towards a particular music genre. This finding that network communities are not composed of users that listen to the same music 
is confirmed in another study on Last.fm~\cite{guidotti2019know}. Furthermore, in the same study~\cite{guidotti2019know}, Guidotti and Rossetti's results show that users who listen to various genres tend to connect with people with high music preference entropy (a measure of diversity), whereas users who listen to music comprising few genres tend to connect with users with a narrow music taste. They also conclude that Last.fm users tend to cluster with peers with similar music entropy and/or similar temporal listening behaviors, computed using frequency dictionaries based on both day and time of the day. Another study tackling the topic of diversity and social ties on the music platforms \textit{Netease Music} and \textit{Weibo} demonstrates that users with low diversity are more similar in terms of music taste 
and that it is difficult for high diversity users to find friends sharing similar music preferences~\cite{zhou2017homophily}. However, to the best of our knowledge, we are among the first to investigate homophily and link prediction as an interplay between mainstreaminess, novelty, diversity, and listening profile similarity in music preferences.
\section{Data and preprocessing}\label{data-and-preprocessing}

We use the well-established LFM-1b~\cite{schedl2016lfm} and LFM-1b UGP~\cite{schedl2017large} datasets\footnote{\url{http://www.cp.jku.at/datasets/LFM-1b/}} which include the following user data of our interest: (i) user-artist playcount matrix, (ii) user-genre playcount matrix, and (iii) features describing user preferences towards mainstream, novel, and diverse content~\cite{schedl2016lfm}. 

LFM-1b contains more than one billion listening events (LEs) of $120,\!175$ unique users from the music platform Last.fm. 
Furthermore, Last.fm allows users to connect with others on the platform by establishing friendship connections. 
A list of each user's friends is available through the Last.fm API\footnote{\url{https://www.last.fm/api/show/user.getFriends}}. 
To create a representative social (friendship) network for our study, we adopt the breadth-first-search sampling strategy with LFM-1b users as seed nodes and crawl friends up to two hops away.
For users that are not part of the LFM-1b dataset, no information about M, N, and D are available. 
Therefore, we keep only those edges where both nodes connected by the edge are part of the original LFM-1b dataset. This results in a network, i.e., an undirected unweighted graph $G=(V,E)$, consisting of $|V|=11,\!792$ nodes (users) and $|E|=78,\!989$ edges (friendship connections).

We model users' listening profiles from (i) and (ii) (on the artist and genre level) and summarize (iii) in the following.

\subsection{Listening profiles}\label{subsec:listening-profiles}

We create an artist and a genre profile for each user. A single occurrence of a user listening to a track is called a listening event (LE).
From the LEs, we construct the user-artist profile matrix $\matr{L}^{artist}$ based on user-artist playcounts\footnote{Number of listening events associated with a user $u$ and an artist $i$.}, and a user-genre profile matrix $\matr{L}^{genre}$ based on user-genre playcounts\footnote{Number of listening events associated with a user $u$ and a genre $j$.}.  
Each value $L^{artist}_{u,i}$ represents the number of times user $u$ has listened to an artist $i$, whereas $L^{genre}_{u,j}$ stands for number of times user $u$ has listened to a genre~$j$. We create a genre profile $L^{genre}_{u,*}$ for each user $u$ as a $k$-dimensional vector, using $k=n_{genre}=1,\!998$ genres and styles from Freebase\footnote{\url{https://developers.google.com/freebase}}. Each genre profile vector is computed based on the user-artist interactions, where each artist is described as a weighted bag-of-words representation of genres. We use the variant weighted by playcount from~\cite{schedl2017large}. Additionally, each row vector in both listening profile matrices is normalized so that its Euclidean norm equals $1$. As a result, each normalized vector $L^{artist}_{u,*}$ represents an artist profile of user $u$, and $L^{genre}_{u,*}$ represents $u$'s genre profile, as described in~\cite{schedl2015tailoring}. However, 
$\matr{L}^{artist}$ is very sparse (density is $0.087\%$) as many users only listen to a very small fraction of all 
artists in the dataset. 
To address this sparsity issue, we apply non-negative matrix factorization (NMF)\footnote{We used the implementation from \url{https://github.com/kimjingu/nonnegfac-python}} to $\matr{L}^{artist}$.
NMF is a widely used tool for the analysis of high-dimensional data as it automatically extracts sparse and meaningful features from a set of non-negative data vectors~\cite{gillis2020nonnegative}. We empirically determined the number of reduced dimensions as 20, and refer to the transformed representation of $\matr{L}^{artist}$ 
as $\matr{W}^{artist}$. In the remainder of the paper, we use artist profiles for further analyses and genre profiles to compute diversity features.

\subsection{User mainstreaminess (M), novelty (N), and diversity (D) 
}\label{subsec:mnd-features}


In this section, we describe how M, N, and D are computed. We explore a number of features from the literature and also propose a diversity measure of our own 
to complement existing formulations which yield results of varying quality, both in our experiments but also 
studies on recommender systems~\cite{schedl2015tailoring, bauer2019global, schedl2017distance}. Besides, we explain how we divide users into groups based on the resulting feature values. 

\para{Mainstreaminess (M).} 
The mainstreaminess features we use are provided in the LFM-1b dataset~\cite{schedl2015tailoring}. More specifically, they are calculated as the overlap between the user’s listening history and an aggregated listening history of all users, averaged over time windows of 1, 6, and 12 months ($M_u^{1m}$/$M_U^{6m}$/$M_U^{12m}$), as well as over the entire period of the user's listening activity ($M_u^G$).

\para{Novelty (N).} 
We adopt three features proposed in~\cite{schedl2015tailoring} with values provided in the LFM-1b dataset. According to these features, a user's inclination to listen to novel music is quantified by the percentage of new artists listened to, averaged over time windows of 1, 6, and 12 months ($N_u^{1m}$/$N_u^{6m}$/$N_u^{12m}$).

\para{Diversity (D).} 
The LFM-1b dataset does not include explicit features of diversity other than simple numbers of unique tracks and artists listened to by the user. We log-normalize those counts and denote them as $D_u^{tracks}$ and $D_u^{artists}$, respectively. To add more sophisticated diversity measures, we 
compute genre coverage $D_u^{GC}$ and genre entropy $D_u^{GE}$ using the genre profile $L_{u,*}^{genre}$ as input. $D_u^{GC}$ is computed similarly as tag coverage and tag entropy in~\cite{deldjoo2019retrieving}. More specifically, $D^{GC}$ computes the percentage of all genres listened to by user $u$ among all $1,\!998$ genres from Freebase, 
whereas $D_u^{GE}$ is computed as the entropy of the distribution of genre occurrences of all artists listened to by user $u$. 
Additionally, we propose a novel diversity feature, i.e., the \textit{weighted average genre diversity} $D^{w\_avg}_u$, calculated from the $u$'s genre playcount vector $L^{genre}_{u,*}$ as follows:

\begin{equation}\label{eq:1}
    D^{w\_avg}_u = \frac{\sum^{n_{genre}}_{i=1} \frac{L_{u,i}}{\mathbf{max} \hspace{0.25em} (L_{u,*})}}{n_{genre}} 
\end{equation}

\para{User groups.} For each M, N, and D feature, we categorize users into a \textit{low}, \textit{mid}, or \textit{high} group based on the corresponding feature value. Decision thresholds are calculated as in~\cite{schedl2015tailoring}. More specifically, in the initial step, users' feature values are first sorted in ascending order. We proceed by summing the user feature values from the beginning until we reach one third of the total sum, and assign all those users that contributed to the sum into the low value group. Users are assigned similarly into the medium value group as we continue summing the values until two thirds are reached. The remaining users are assigned to the high value group. These group assignments are utilized later in the paper when we study observed to expected edge ratio as well as features in link prediction experiments.

\subsection{Preparing the dataset for link prediction}\label{subsec:link-prediction-dataset}

By approaching link prediction as a binary classification problem, we need to prepare the dataset so that a model can be applied to predict a class label $y \in \left\{ 0,1 \right\}$ given a feature vector $\mathbf{x}$.
Observed edges are positive class instances, while missing edges are treated as negative class instances. Here is where the extreme class imbalance issue arises, as there are in total $|E|$ positive classes and $\frac{|V| \times |V|}{2} - |E|$ negative classes. Therefore, we resort to negative class sampling, i.e., we randomly sample $|E|$ negative class instances (missing edges) so that the resulting dataset consists of $2 \times |E|$ instances. 
This purely random sampling of negative classes is the most appropriate sampling method according to the link prediction evaluation guidelines in~\cite{yang2015evaluating}. Furthermore, according to the same guidelines, evaluation results are stable when the sampled negative class percentage is between $10^{-1}$ and $10^2$, and in our case, $|E|$ sampled negative class instances amount to $0.11 \%$. To further account for randomness and make our results more robust, we create 10 datasets in the same manner with different random seeds. In our experiments, we average the results on all 10 datasets.

Each edge instance between a node (user) pair $(u,v)$ is described with a class label $y \in \left\{0,1\right\}$ and a feature vector $\mathbf{x} = \mathbf{x_u}^\frown \mathbf{x_v}^\frown \mathbf{x_\Delta}$ where $\mathbf{x_u}$ and $\mathbf{x_v}$ represent feature vectors of nodes $u$ and $v$, respectively, and $\mathbf{x_\Delta} = f(\mathbf{x_u}, \mathbf{x_v})$ represents features derived from both users. More detailed descriptions of feature vector $\mathbf{x}$ can be found in Table~\ref{tab:feature-vector-description}. We can categorize features into: 
(i) M, N, and D features (MNDF), (ii) artist profile features (APF), and (iii) graph-based features (GF). We utilize graph-based features commonly used for link prediction (number of common neighbors, Jaccard index, and Adamic-Adar index)~\cite{aggarwal2015data} in order to provide perspective to the merit of (i) and (ii) by comparing them to an "unfairly" strong baseline~\cite{bischoff2012we}. 

\begin{table}[h!]
\centering
\caption{\label{tab:feature-vector-description} \textbf{Feature vector $\mathbf{x}$ used in link prediction experiments is a result of concatenating features in this table. Features are binary, categorical, or numeric. Relative difference in $\mathbf{x_\Delta^M}$, $\mathbf{x_\Delta^N}$, and $\mathbf{x_\Delta^D}$ is calculated as $\left|\frac{a-b}{\frac{a+b}{2}}\right|$}.}
\hspace{-2.5mm}\scalebox{0.67}{
\begin{tabular}{?c|c|c?}
\Xhline{2\arrayrulewidth}
\textbf{Notation} & \textbf{Description} & \textbf{Type} \\ \Xhline{2\arrayrulewidth}
    \multicolumn{3}{?c?}{\textit{Mainstreaminess, novelty, and diversity features (MNDF)}} \\ \Xhline{2\arrayrulewidth}
    $\mathbf{x^M_{\left\{u,v\right\}}}$     &  4 mainstreaminess values per user           & Numeric vector      \\ \hline 
    $\mathbf{x^N_{\left\{u,v\right\}}}$     &  3 novelty values per user           & Numeric vector     \\ \hline
    $\mathbf{x^D_{\left\{u,v\right\}}}$     &  5 diversity values per user           & Numeric vector     \\ \hline
    \multirow{2}{*}{$\mathbf{x^{M_{group}}_{\left\{u,v\right\}}}$} & Low/mid/high group categorization for each   & \multirow{2}{*}{Categorical vector} \\ 
    &  mainstreaminess feature and each user & \\ \hline
    \multirow{2}{*}{$\mathbf{x^{N_{group}}_{\left\{u,v\right\}}}$} & Low/mid/high group categorization   & \multirow{2}{*}{Categorical vector} \\ 
    & for each novelty feature and each user & \\ \hline
    \multirow{2}{*}{$\mathbf{x^{D_{group}}_{\left\{u,v\right\}}}$} & Low/mid/high group categorization  & \multirow{2}{*}{Categorical vector} \\ 
    & for each diversity feature and each user & \\ \hline
    \multirow{2}{*}{$\mathbf{x_\Delta^M}$} & Relative difference between respective & \multirow{2}{*}{Numeric vector} \\ 
    & user mainstreaminess features   & \\ \hline
    \multirow{2}{*}{$\mathbf{x_\Delta^N}$} & Relative difference between respective & \multirow{2}{*}{Numeric vector} \\ 
    & user novelty features & \\ \hline
    \multirow{2}{*}{$\mathbf{x_\Delta^D}$} & Relative difference between respective & \multirow{2}{*}{Numeric vector} \\ 
    & user diversity features  & \\ \hline
    \multirow{2}{*}{$\mathbf{x_\Delta^{M_{group}}}$} & True if both users are categorized in the  & \multirow{2}{*}{Binary vector} \\ 
    & same mainstreaminess group, False otherwise & \\ \hline
    \multirow{2}{*}{$\mathbf{x_\Delta^{N_{group}}}$} & True if both users are categorized in the  & \multirow{2}{*}{Binary vector} \\ 
    & same novelty group, False otherwise & \\ \hline
    \multirow{2}{*}{$\mathbf{x_\Delta^{D_{group}}}$} & True if both users are categorized in the  & \multirow{2}{*}{Binary vector} \\ 
    & same diversity group, False otherwise & \\ \Xhline{2\arrayrulewidth}
    \multicolumn{3}{?c?}{\textit{Artist profile features (APF)}} \\ \Xhline{2\arrayrulewidth}
    $\mathbf{x}_{\left\{u,v\right\}}^{\mathbf{W^{artist}}}$     & Low-dimensional user artist profile vectors & Numeric vector     \\ \hline
    $x_\Delta^{\mathbf{W^{artist}}}$     & Cosine similarity between user artist profile vectors & Numeric scalar     \\ \Xhline{2\arrayrulewidth}
    \multicolumn{3}{?c?}{\textit{Graph-based features (GF)}} \\ \Xhline{2\arrayrulewidth}
    $x_\Delta^{CN}$     & Number of common neighbors between users & Numeric scalar     \\ \hline
    $x_\Delta^{J}$     & Jaccard index between users & Numeric scalar     \\ \hline
    $x_\Delta^{AA}$     & Adamic-Adar index between users & Numeric scalar     \\ \Xhline{2\arrayrulewidth}
\end{tabular}
}
\end{table}
\section{Homophily in User Preferences}

In this section, we describe how we compute and present our results on homophily based on users' artist profiles and mainstreaminess, novelty, and diversity.  



\subsection{Artist profile homophily}

We investigate if there is homophily in the Last.fm friendship graph $G$ based on users' artist profiles. 
To that end, we compare similarities between artist profile vectors of connected users with similarities of randomly connected users. 

\begin{figure}[h!]
    \centering
    \includegraphics[width=0.35\textwidth]{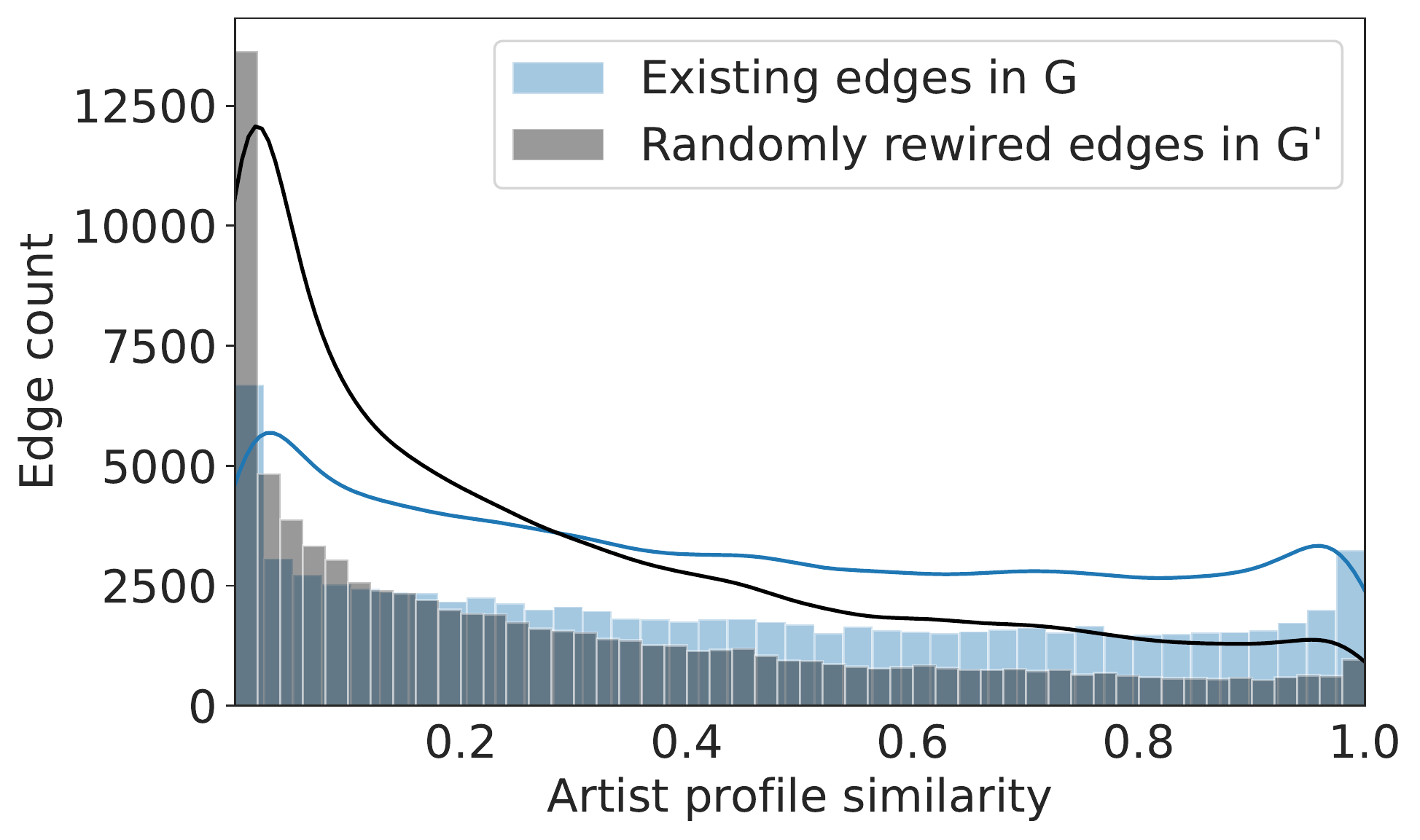}
    \caption{\textbf{Artist profile similarities calculated as dot products of user artist profile vectors from $\matr{W}^{artist}$ between connected nodes in graph $G$ and connected node pairs in $G'$ created by randomly rewiring edges in $G$ while preserving the degree distribution and homophily of M, N, and D. The visualization depicts a histogram plot with a corresponding kernel density estimate (KDE) plot.}}
    \label{fig:listening-profile-similarity}
\end{figure}

We first calculate the dot product between the artist profile vectors (from $\matr{W^{artist}}$) of the connected users. Then, we create a baseline random graph $G'$ for comparison. We select an adapted version of the configuration model from network science~\cite{newman2003structure} to preserve two graph properties of interest, namely the node degree distribution and homophily for M, N, and D.
In this random graph model, each edge of a user $u$ is randomly rewired so that if $u$ was connected to a user $v$ belonging to low M, medium N, and low D user groups, a new edge connects it to a random user $v'$ from the same user groups while also preserving the same node degree as in the original graph. In this way, we ensure that the random graph $G'$ preserves two above-mentioned important properties of the original graph $G$. We report the rewiring probabilities of this configuration model in our GitHub repository.
Finally, we calculate dot product on artist profiles of connected node pairs in the resulting graph $G'$ and compare the two distributions. Our results show that according to the Mann-Whitney U test ($p<0.001$), there is a significant difference between the two distributions as we can observe in Figure~\ref{fig:listening-profile-similarity}. Connected node pairs in $G$ have significantly higher similarity between artist profiles ($mean=0.44$) as opposed to connected node pairs in $G'$ ($mean=0.28$). We can conclude that we observe notable homophily based on user artist profiles in the Last.fm network.

\subsection{Homophily based on M, N, and D}

Next, we quantify homophily in the network for M, N, and D, which are all numeric features. A simple method used to quantify homophily based on numeric features is the \textit{numeric assortativity coefficient $r$}~\cite{newman2018networks}. It is calculated as the Pearson correlation coefficient computed between connected node pairs for a particular node feature.
Assortativity coefficient $r$ values range from $-1$ (completely heterophilic network) to $1$ (completely homophilic network). We calculate $r$ for each M, N, and D feature reported in Section~\ref{subsec:mnd-features} and report the results in Table~\ref{tab:assortativity-results}.

\begin{table}[h!]
\centering
\caption{\label{tab:assortativity-results} \textbf{Assortativity coefficients $r$ for mainstreaminess (M), novelty (N), and diversity (D) vary depending on how their scores are calculated. Values in bold denote the scores with the highest $\mathbf{r}$ for M, N, and D.}}
\scalebox{0.7}{
\begin{tabular}{cc?c?}
\cline{3-3} \cline{3-3}
                                        &  & \textbf{r}     \\ \Xhline{2\arrayrulewidth}
\multicolumn{1}{|c?}{\multirow{4}{*}{\textbf{M}}} & $M^{1m}$ & 0.055          \\ \cline{2-3} 
\multicolumn{1}{|c?}{}                  & $M^{6m}$ & 0.050          \\ \cline{2-3} 
\multicolumn{1}{|c?}{}                  & $M^{12m}$ & 0.043          \\ \cline{2-3} 
\multicolumn{1}{|c?}{}                  & $M^{G}$ & \textbf{0.104} \\ \hline \hline
\multicolumn{1}{|c?}{\multirow{3}{*}{\textbf{N}}} & $N^{1m}$ & 0.063          \\ \cline{2-3} 
\multicolumn{1}{|c?}{}                  & $N^{6m}$ & \textbf{0.111} \\ \cline{2-3} 
\multicolumn{1}{|c?}{}                  & $N^{12m}$ & 0.058          \\ \hline \hline
\multicolumn{1}{|c?}{\multirow{5}{*}{\textbf{D}}} & $D^{tracks}$ & 0.076          \\ \cline{2-3} 
\multicolumn{1}{|c?}{}                  & $D^{artists}$ & 0.104          \\ \cline{2-3} 
\multicolumn{1}{|c?}{}                  & $D^{GC}$ & 0.148          \\ \cline{2-3} 
\multicolumn{1}{|c?}{}                  & $D^{GE}$ & 0.140          \\ \cline{2-3} 
\multicolumn{1}{|c?}{}                  & $D^{w\_avg}$ & \textbf{0.227} \\ \Xhline{2\arrayrulewidth}
\end{tabular}
}
\vspace{-4mm}
\end{table}

We can observe that the Last.fm social network exhibits positive assortativity coefficients for M, N, and D. Overall, it is most pronounced for D (as reflected in the highest correlation coefficients $r$). 
For mainstreaminess, the highest homophily can be observed based on the $M^G$ score, $N^{6m}$ for novelty, and $D^{w\_avg}$ for diversity. 
Thus for categorizing users into groups and the following analyses, we use $M^G$, $N^{6m}$, and $D^{w\_avg}$  
 as representative features of users' mainstreaminess, novelty, and diversity and refer to them from this point in the text simply as M, N, and D. 

\subsection{Between- and within-group observed to expected edge ratio}\label{subsec:between-within}

\begin{figure*}[h!]
\centering
\subfloat[][O/E ratio in $\mathbf{G}$]{
\includegraphics[width=0.24\textwidth]{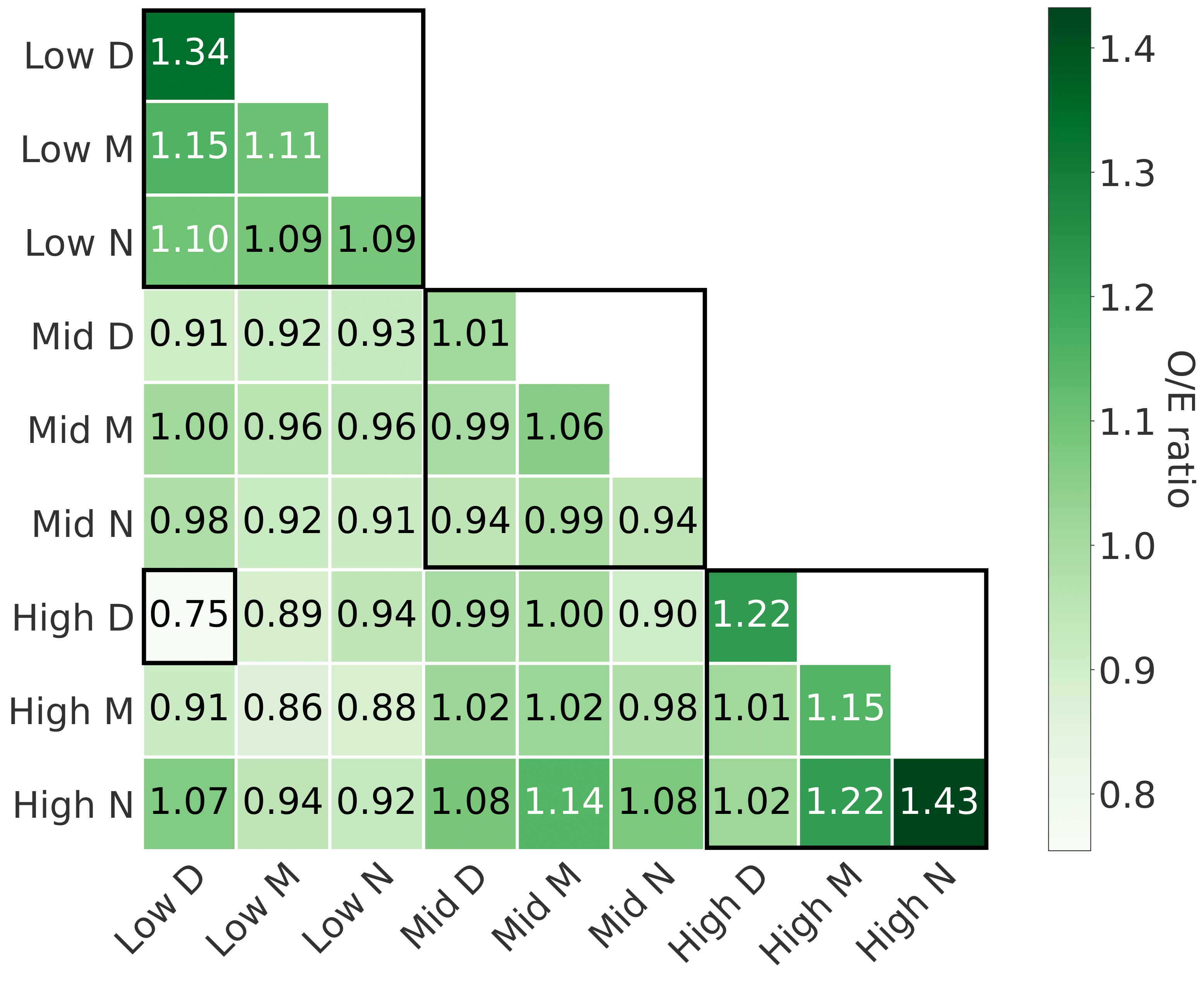}
\label{fig3:subfig1}}
\subfloat[][Artist profile similarity in $\mathbf{G}$]{
\includegraphics[width=0.24\textwidth]{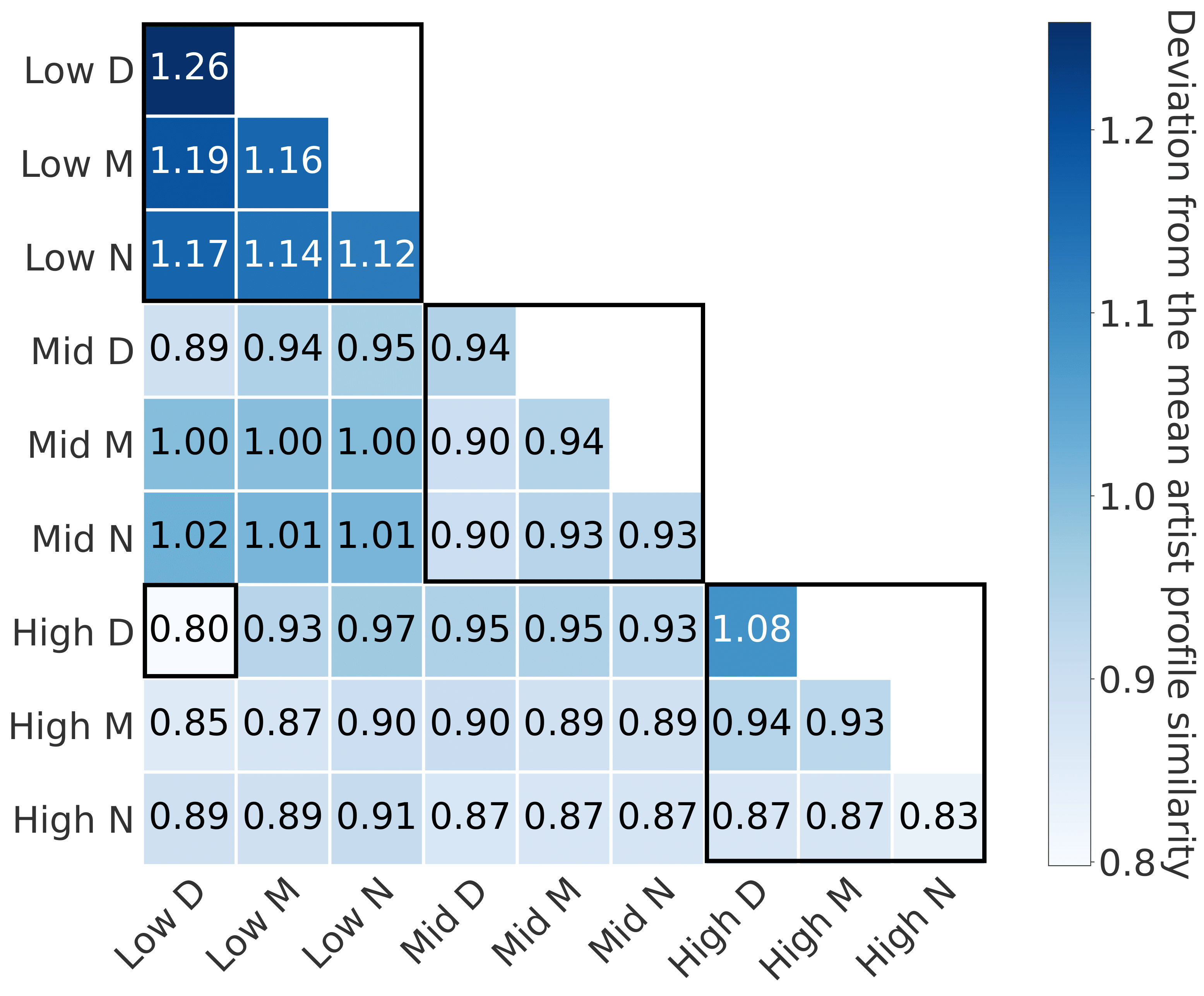}
\label{fig3:subfig2}}
\subfloat[][O/E ratio in $\mathbf{G'}$]{
\includegraphics[width=0.24\textwidth]{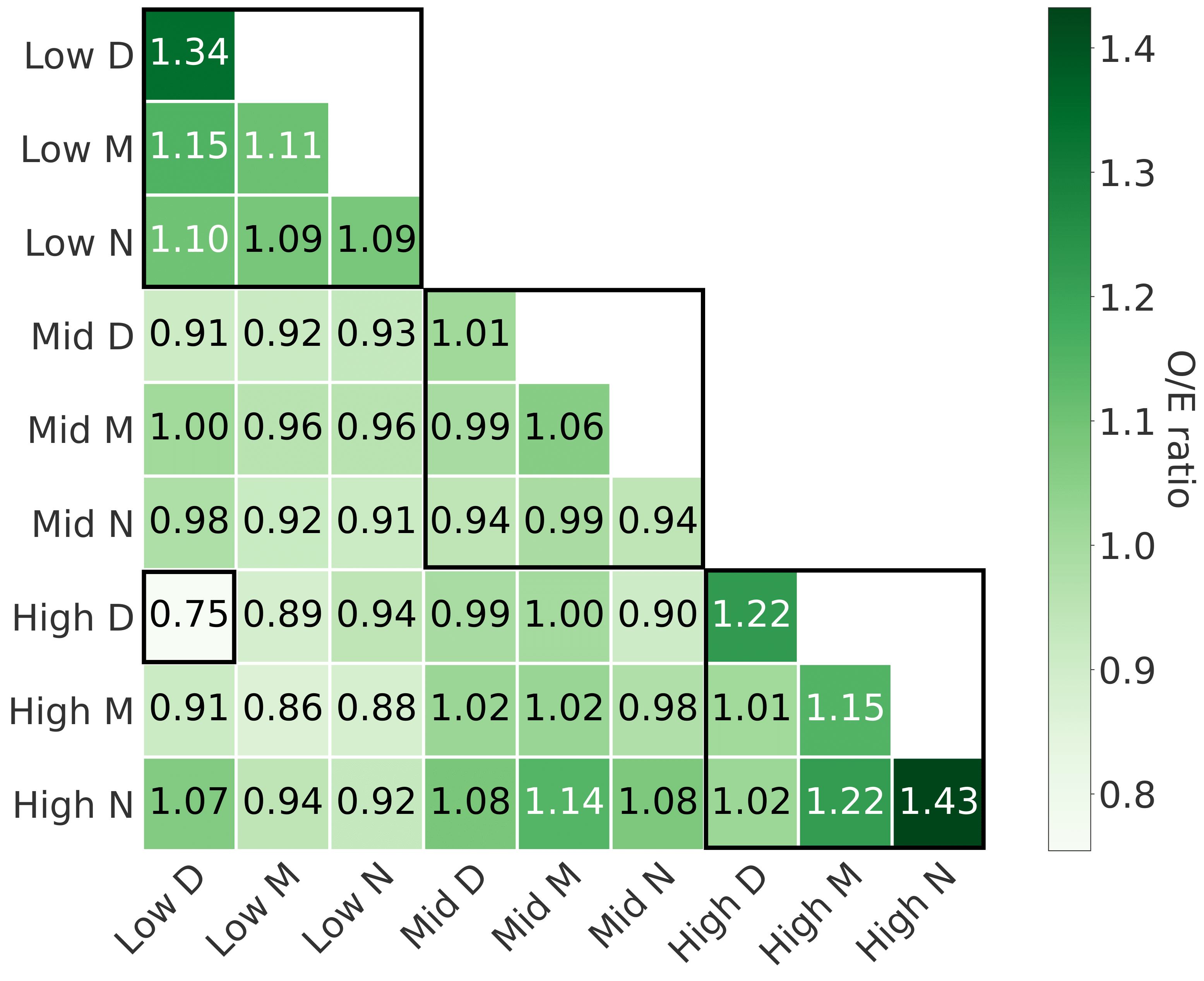}
\label{fig3:subfig3}}
\subfloat[][Artist profile similarity in $\mathbf{G'}$]{
\includegraphics[width=0.24\textwidth]{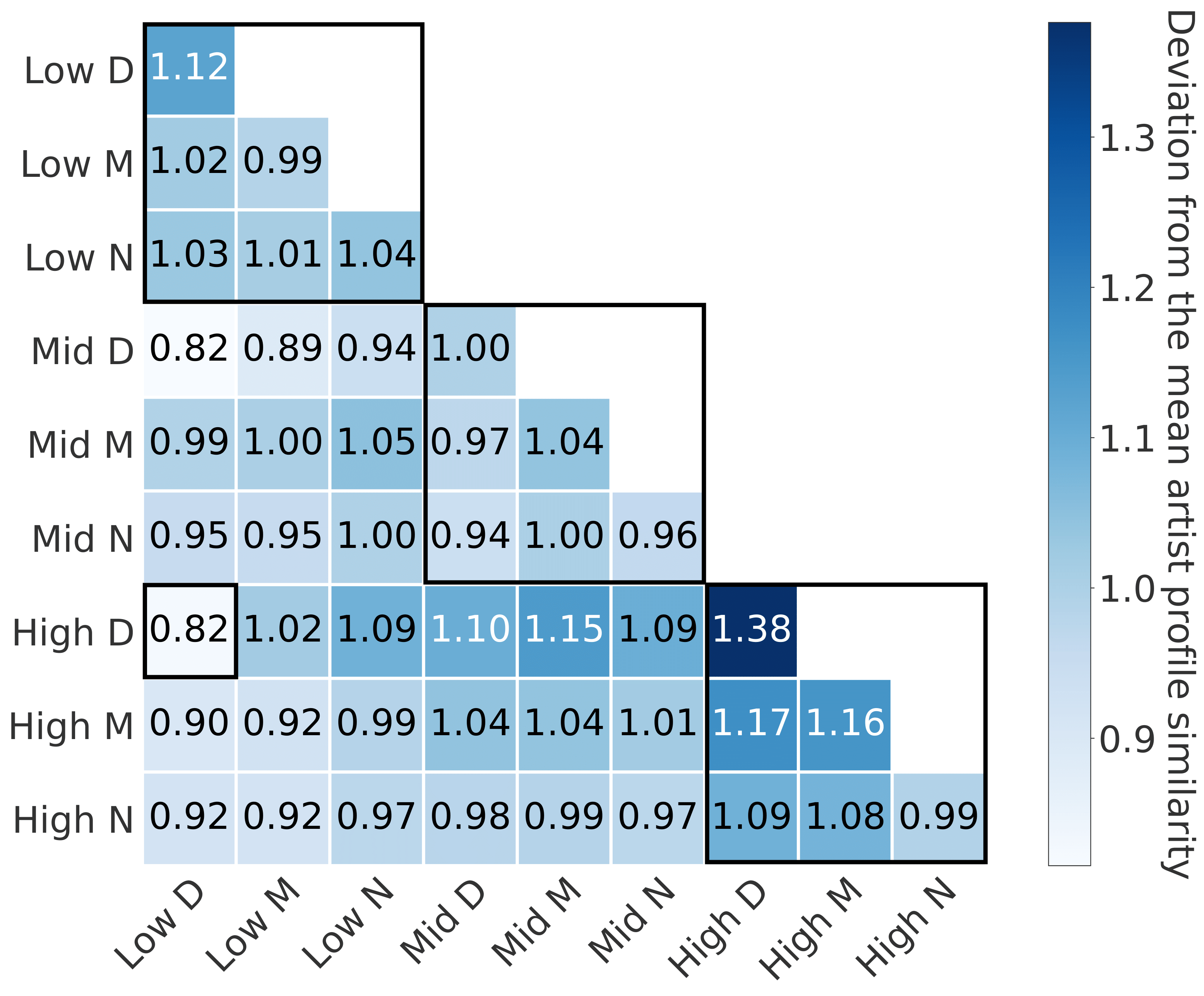}
\label{fig3:subfig4}}
\caption{\label{fig3:globfig} \textbf{Observed/expected (O/E) edge ratio in $\mathbf{G}$ (a) and $\mathbf{G'}$ (c) for between- and within-group edges based on low/medium/high values of M, N, and D. Randomly generated graph $\mathbf{G'}$ preserves the number of between- and within-group edges resulting in almost equal O/E ratio as in the original graph $\mathbf{G}$. Artist profile similarity depicted in Figures (b) and (d) is calculated as the ratio between the average value over the set of between- or within-group edges and the average value over all edges. To account for randomness, values in (c) and (d) are averaged over 10 different random graphs. In each figure, diagonal cells represent values for within-group edges and non-diagonal cells represent values for between-group edges. In the discussion of our results, we focus on values highlighted in black rectangles. }}
\end{figure*}

Assortativity coefficients can only provide us with information on the overall assortativity patterns in the network. However, what often happens is that homophily exists only on some intervals of the feature distribution, e.g., for users with particularly high or low values. 
Referring to the low/mid/high user groups defined in  Section~\ref{subsec:mnd-features}, we investigate 
whether there is homophily in these user groups by counting the observed edges within each group and dividing it with the number of expected edges, i.e., the number of between- and within-group edges assuming fully random pairing with the same amount of nodes and edges as in $G$ (as discussed in~\cite{easley2012networks}). If this ratio is higher than 1, then this points to homophily within a particular user group. By applying the same approach for edges between two different groups, we can determine if there is an unusually higher or lower edge count for users from these two groups. 
In a random graph, the expected number of edges within a particular group (e.g., low D) is $p^2d(G)$ and between two different groups (e.g., between low D and high D) $2 p q d(G)$, where $p$ is the number of nodes in the first group, $q$ is the number of nodes in the second group, and $d(G)$ is the density of the graph $G$. The results of this analysis can be found in Figure~\ref{fig3:subfig1}. When it comes to users' artist profiles, one would expect that groups where homophily is more pronounced also exhibit higher than average artist profile similarity (as similarity breeds connection). Therefore, we show how much does the mean artist profile similarity between two groups deviate from the overall mean artist profile similarity in Figure~\ref{fig3:subfig2} as well as what is expected regarding the artist profile similarity in a random graph $G'$ (Figure~\ref{fig3:subfig4}). To account for randomness, Figures~\ref{fig3:subfig3} and \ref{fig3:subfig4} depict average results over 10 graphs created using different random seeds. Notice that the O/E ratios for user groups are almost identical in $G$ and $G'$ (Figures~\ref{fig3:subfig1} and \ref{fig3:subfig3}) confirming that the synthetic random graph $G'$ preserves homophily based on M, N, and D.




By looking at the values on the diagonal in Figure~\ref{fig3:subfig1}, we can observe that all groups except for the medium N group exhibit homophilic behavior. This behavior is most strongly pronounced in the high N group, but also strong in the high M, low D, and high D user groups. Low value groups also have notably higher artist profile similarity when compared with the random graph $G'$. 
However, when it comes to high value groups, they have lower similarity when compared with $G'$ and strong homophily values, especially for high N and high D users. These results could suggest that \textit{high-novelty-seekers} or \textit{explorers} connect with other high N users that help them discover new music content. This behavior also holds for high D users, but to a lesser extent. 

When it comes to between-group edges, we would like to highlight that the high D group has a notably smaller number of edges with the low D group which is likely due to them having a lower similarity in the artist profiles (see Figure~\ref{fig3:subfig2}) and is an example where dissimilarity breeds disconnection as this behavior also appears in the random graph $G'$. 

\vspace{3mm}

\noindent\fbox{%
    \parbox{0.48\textwidth}{%
        \textbf{Finding 1:} In a comparison with a random graph model, our results show that users that connect are significantly more similar with respect to their artist profiles. Furthermore, there is an observable homophilic behavior for M, N, and D, most pronounced for D. We also discover different levels of homophily in low/mid/high value groups. High N and low D users form the most connections within their groups with an interesting difference, i.e., low D users also have similar artist profiles whereas the similarity is notably lower for high N users. This means that high N users form a large amount of their connections with other high N users even though their artist profiles are not as similar.
    }%
}


\section{Link Prediction With User Preferences}\label{sec:link-prediction}

Since the results of our analysis have shown notable homophily in user mainstreaminess, novelty, and diversity, as well as user artist profile similarity, we now explore their efficacy for the task of link prediction with a supervised learning approach, i.e., binary classification. Approaching link prediction as a binary classification task is challenging due to extreme class imbalance and inconsistent evaluation metrics. Therefore, we design our evaluation experiments according to guidelines proposed by Yang et al.~\cite{yang2015evaluating}. 


We use the dataset as defined in Section~\ref{subsec:link-prediction-dataset} and consider the following objectives in our link prediction experiments. First, we do not focus on the performance comparison of different classification algorithms, but rather on using an established classification algorithm as a tool to explore the merit of M, N, and D in comparison with user artist profiles for the task of link prediction. We provide context to those results by comparing them with a strong baseline using graph-based features as well as a weak random baseline. Additionally, we explore differences in results for different user groups as defined in Section~\ref{subsec:mnd-features}, namely low/mid/high M, N, and D. Finally, we evaluate the contribution of individual features by analyzing feature importance scores in the predictive models.

For these purposes, we use the \textit{Extreme Gradient Boosting} or \textit{XGBoost}\footnote{We use the Python XGBoost implementation from \url{https://xgboost.readthedocs.io/en/latest/python/}}~\cite{chen2016xgboost} for binary classification, which is a type of gradient boosted trees algorithm which has proven to be highly effective for link prediction~\cite{behera2021follower}.  

We compare different combinations of (i) mainstreaminess, novelty, diversity features (\textbf{MNDF}), (ii) artist profile features (\textbf{APF}), and (iii) graph-based features (\textbf{GF}). 
We compare our results with a stratified random classifier as a weak baseline. 

We split our data into train and test subsets using an 80/20 split, and we evaluate our models on 10 different random splits so that each feature combination is evaluated on 10 datasets (with different randomly selected negative samples as described in Section ~\ref{subsec:link-prediction-dataset}) and 10 splits (using different random seeds), which amounts to 100 experiments per feature combination. We do not conduct hyperparameter optimization and use default XGBoost parameters since we are only interested in assessing (relative) performance differences between using different feature combinations and not in outperforming any state-of-the-art approaches. Since we use negative class sampling and evaluate results on balanced datasets (with a 50-50 positive-to-negative class ratio), in order to compare the results, we use the F1 score as the performance metric.

Finally, XGBoost provides feature importance scores out-of-the-box, which indicate how useful or valuable each feature is in the construction of the boosted decision trees within the model. The more a feature is used to make key decisions with decision trees, the higher its relative importance.

\subsection{Accuracy results}

\begin{figure}[ht!]
\centering
\includegraphics[width=0.3\textwidth]{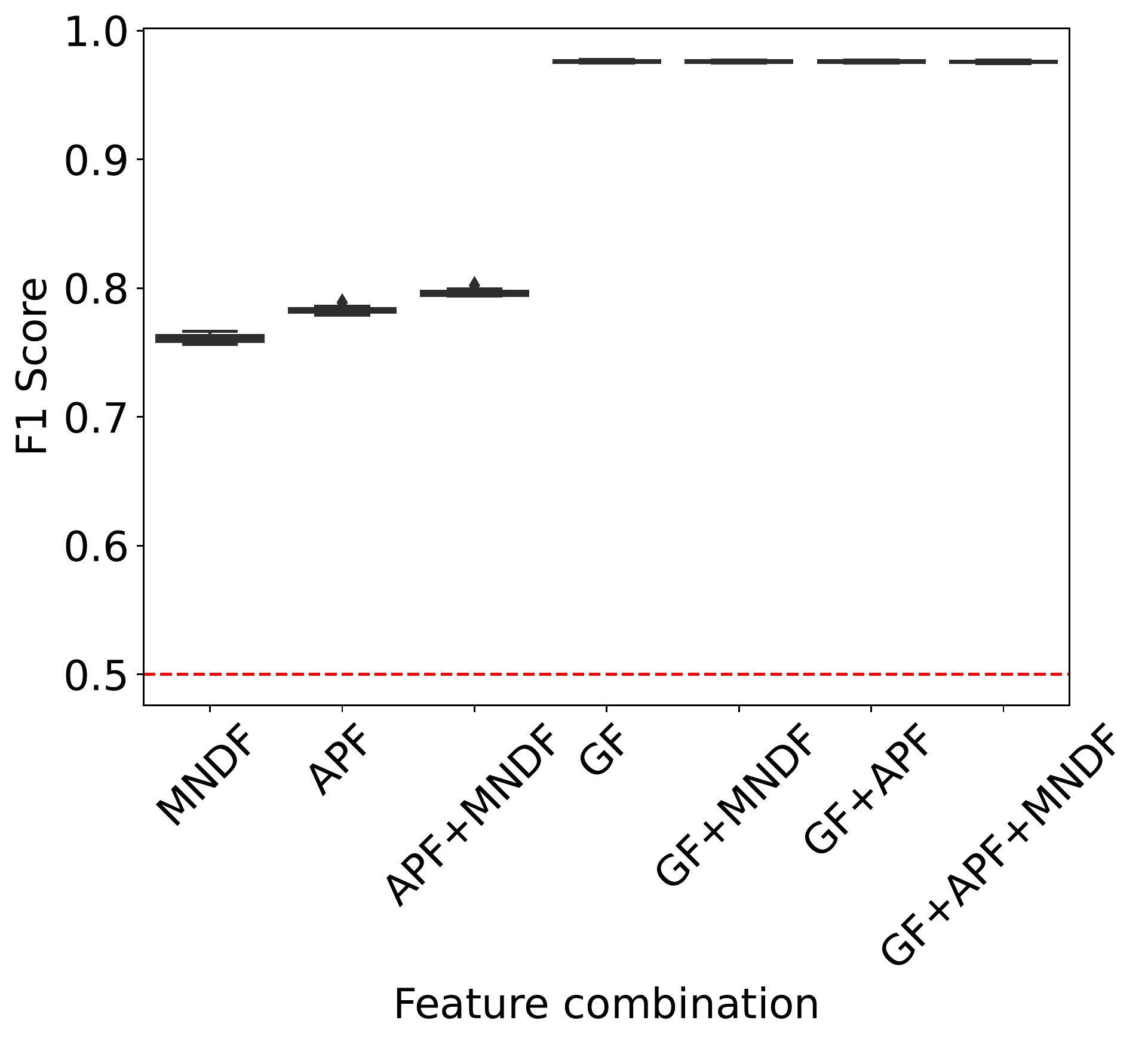}
\caption{\label{fig4:globfig} \textbf{Accuracy results for link prediction using XGBoost measured with F1 score for different feature combinations. For comparison, a stratified random classifier (i.e., weak baseline) yields an F1 score of $0.5$.}}
\vspace{-6mm}
\end{figure}

We present the link prediction results for different feature combinations, namely MNDF, APF, and GF in Figure~\ref{fig4:globfig}. For comparison, F1 score for a stratified random classifier used as a weak baseline is $0.5$. 


The results lead to several interesting insights. First, all feature combinations strongly outperform the weak baseline. Furthermore, using only MNDF results in comparable but slightly worse performance than using only APF which is great considering APF contains significantly more information on user preferences, while a combination of the two shows improvement over using each on their own. Not surprisingly, using GF as a strong baseline firmly outperforms combinations of MNDF and APF because two randomly selected node pairs in a graph will not have many common neighbors as shown in~\cite{bischoff2012we}. Finally, using MNDF or APF in combination with GF does not improve the results in comparison with using only GF. 
Nevertheless, using the MNDF can have an important 
application in a cold-start user recommendation setting in which the user does not yet have many friendship connections. For example, a system could generate meaningful friendship recommendations from user-defined MNDF.

\subsection{Accuracy results for different user groups}

Our next objective is to study differences in link prediction accuracy using MNDF and APF between low/mid/high value user groups based on their M, N, and D values. Aggregated mean F1 scores for each user group are shown in Table~\ref{tab:user-group-link-prediction-results}.

\begin{table}[t!]
\centering
\caption{\label{tab:user-group-link-prediction-results} \textbf{Mean F1 scores aggregated over different user groups based on their M, N, and D values.}}
\scalebox{0.8}{
\begin{tabular}{cc?c|c|c?}
\cline{3-5} \cline{3-5}
                                        &  & \textbf{MNDF} & \textbf{APF} & \textbf{MNDF+APF} \\ \Xhline{2\arrayrulewidth}
\multicolumn{1}{|c?}{\multirow{3}{*}{\textbf{M}}} & Low & 0.7668 & 0.7813 & \textbf{0.7974} \\ \cline{2-5} 
\multicolumn{1}{|c?}{}                  & Medium & 0.7602 & 0.7818 & \textbf{0.7859} \\ \cline{2-5} 
\multicolumn{1}{|c?}{}                  & High & 0.7761 & \textbf{0.7946} & 0.7908 \\ \hline \hline
\multicolumn{1}{|c?}{\multirow{3}{*}{\textbf{N}}} & Low & 0.7655 & 0.7811 & \textbf{0.7953} \\ \cline{2-5} 
\multicolumn{1}{|c?}{}                  & Medium & 0.7643 & 0.7815 & \textbf{0.7872 }\\ \cline{2-5} 
\multicolumn{1}{|c?}{}                  & High & 0.7811 & 0.7958 & \textbf{0.7971} \\ \hline \hline
\multicolumn{1}{|c?}{\multirow{3}{*}{\textbf{D}}} & Low & 0.7671 & 0.7817 & \textbf{0.7971} \\ \cline{2-5} 
\multicolumn{1}{|c?}{}                  & Medium & 0.7602 & 0.7878 & \textbf{0.7908} \\ \cline{2-5} 
\multicolumn{1}{|c?}{}                  & High & 0.7668 & 0.7829 & \textbf{0.7924} \\ \Xhline{2\arrayrulewidth}
\end{tabular}
}
\end{table}

Our results show that using MNDF+APF performs better for all user groups except for the high M group, where using only APF yields the best results. Using only MNDF shows the best results for the high N group and worst for medium M and medium D. Using a combination of MNDF and APF shows the strongest improvement for low value groups.

\subsection{Feature importance}

Furthermore, we explore the merit of using MNDF with APF for link prediction. Thus, we compute XGBoost feature importance scores using the MNDF+APF approach and present the obtained results below for features defined in Table~\ref{tab:feature-vector-description}. For vector-type features, we report the sum of importance scores. Out of single features, the one with the highest importance score is similarity in user artist profiles $x_{\Delta}^{\matr{W_{artist}}}$.


\begin{figure}[h!]
  \begin{minipage}[t]{0.5\linewidth}
    \vspace{0pt}
    \centering%
    \scalebox{0.65}{
    \begin{tabular}{?c|c?c|c?}
    \Xhline{2\arrayrulewidth}
    \textbf{Feature} & \textbf{Importance} & \textbf{Feature} & \textbf{Importance} \\ \Xhline{2\arrayrulewidth}
    $\mathbf{x^M_{\left\{u,v\right\}}}$ & 0.0777 & $\mathbf{x_\Delta^N}$ & 0.0096 \\ \hline
    $\mathbf{x^N_{\left\{u,v\right\}}}$ & 0.0600 & $\mathbf{x_\Delta^D}$ & 0.0180 \\ \hline
    $\mathbf{x^D_{\left\{u,v\right\}}}$ & 0.1001 & $\mathbf{x_\Delta^{M_{group}}}$ & 0.0100 \\ \hline
    $\mathbf{x^{M_{group}}_{\left\{u,v\right\}}}$ & 0.0559 & $\mathbf{x_\Delta^{N_{group}}}$ & 0.0094   \\ \hline
    $\mathbf{x^{N_{group}}_{\left\{u,v\right\}}}$ & 0.0490 & $\mathbf{x_\Delta^{D_{group}}}$ & 0.0157 \\ \hline
    $\mathbf{x^{D_{group}}_{\left\{u,v\right\}}}$ & 0.0726& $\mathbf{x}_{\left\{u,v\right\}}^{\mathbf{W^{artist}}}$ & 0.4130 \\ \hline
    $\mathbf{x_\Delta^M}$ & 0.0124 & $x_\Delta^{\mathbf{W^{artist}}}$ & 0.0964 \\ \Xhline{2\arrayrulewidth}
    \end{tabular}
    }
    \par\vspace{0pt}
  \end{minipage}
  \hspace{5mm}
  \begin{minipage}[t]{0.42\linewidth}
    \raggedright
    \vspace{0pt}
    \includegraphics[width=\textwidth]{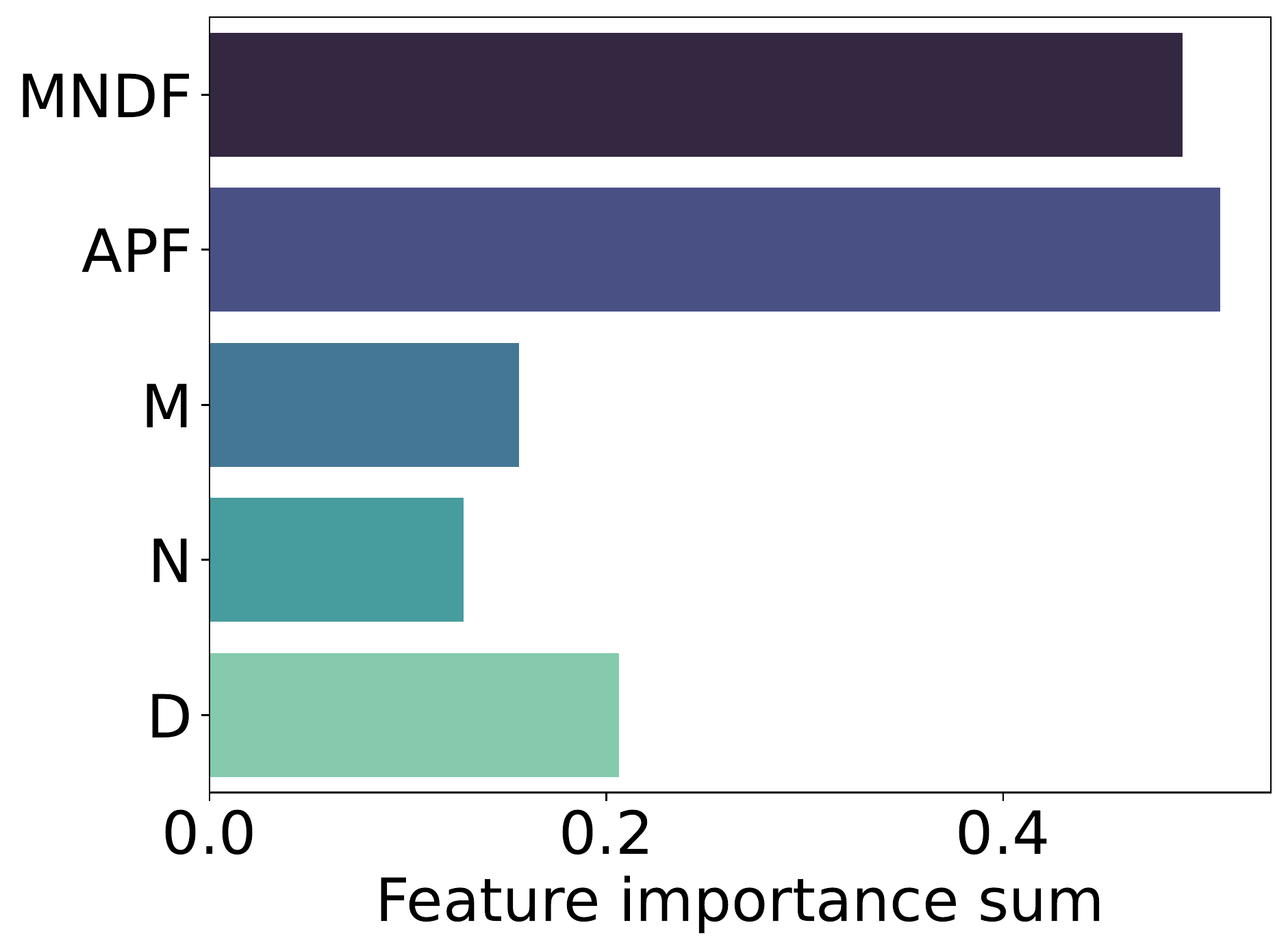}
    \par\vspace{0pt}
  \end{minipage}%
\caption{\textbf{Table (left) depicts aggregated feature importance scores given as a result from the MNDF+APF approach grouped by categories of feature vector $\mathbf{x}$ as defined in Table~\ref{tab:feature-vector-description}. Figure (right) shows aggregated feature importance scores differently to highlight differences between MNDF, APF, M, N, and D.}}
\label{fig:feature-importances}
\end{figure}

Finally, we aggregate the feature importance scores into categories, namely MNDF, APF, M, N, and D and show sums of the feature importance scores in Figure~\ref{fig:feature-importances}. 
We observe that MNDF and APF contribute almost equally to the final results of link prediction, and D has more impact than M or N.

\vspace{2mm}

\noindent\fbox{%
    \parbox{0.48\textwidth}{%
        \textbf{Finding 2:} Using M, N, and D features (MNDF) for link prediction outperforms the random baseline and achieves similar results to those achieved using artist profile features (APF), whereas the combination of the two results in best performance. Both MNDF and APF are not able to outperform 
        a strong baseline using graph features (GF) for link prediction. Nevertheless, they can be useful in cold-start scenarios for user recommendation. 
        Furthermore, we find notable differences in link prediction accuracy for different low/mid/high user groups. Finally, looking at feature importance scores in the MNDF+APF approach, we observe the equal contribution of MNDF and APF with the most important single feature being the artist profile similarity between users.
    }%
}

\section{Conclusion and discussion}

In this paper, we study homophily and link prediction in the online social network of the music platform Last.fm. More specifically, we explore homophily concerning users' music preferences using 
their artist profiles as well as their inclination towards listening to mainstream, novel, and diverse content. We confirm the existence of homophily for users' artist profiles, showing that users that are friends online are more similar with respect to artists they listen to than a random pair of users. Furthermore, we show that there exists homophily concerning features describing users' mainstreaminess (M), novelty (N), and diversity (D) with the strongest exhibited homophily for~D. 
Looking into low/mid/high user groups with respect to M, N, and D, we observe different behaviors: 
First, users from certain groups, like the low D group, form many connections with other low D users. These connections correlate with the artists they listen to, as they have high similarities in their artist profiles. However, the causality is unclear here as in whether they have more connections among them due to homophily because they listen to similar music or if it is the other way around, and they listen to similar music due to underlying effects of social contagion~\cite{hodas2014simple}. The limiting factor in trying to answer this question is the available data, as the Last.fm API does not provide temporal information on when the friendship connection was established. Second, we notice another interesting behavior, i.e., high N users form notably more friendship connections with other high N users. However, those user pairs have on average noticeably lower artist profile similarity. One possible explanation for this behavior is that those users are  
explorers, and they connect with other users who enable them to discover new music content. Another explanation could be provided in the social identity theory (SIT)~\cite{tajfel1978differentiation}, which states that groups that people belong to are an important source of pride and self-esteem giving people a sense of social identity. If this was the case, it would mean that high N users identify themselves as such and for this reason seek to connect with other users which are similar to them in the sense of preference towards discovering novel music content. A user study would be needed to test these hypotheses. We conclude the paper with a demonstration of the usefulness of M, N, and D features in a link prediction scenario where we show that they are almost equally useful as user artist profiles and are able in some cases to slightly improve the link prediction accuracy. The applicability of our findings could prove useful in a cold-start user recommendation scenario where users have few or no friendship connections (e.g.,~\cite{duricic2018trust}).

Potential avenues for future work include: (i) using the more fine-grained Spotify genre annotations\footnote{\url{https://developer.spotify.com/documentation/web-api/reference-beta/\#endpoint-get-an-artist}} instead of those from Freebase, (ii) considering more advanced features describing user mainstreaminess~\cite{bauer2019global}, and (iii) studying demographic features available in the dataset. 

\newpage

\para{Acknowledgments.} 
This work is supported by the H2020 project TRUSTS (GA: 871481), the Austrian Science Fund (FWF): P33526, and the “DDAI” COMET Module within the COMET – Competence Centers for Excellent Technologies Programme, funded by the Austrian Federal Ministry for Transport, Innovation and Technology (bmvit), the Austrian Federal Ministry for Digital and Economic Affairs (bmdw), the Austrian Research Promotion Agency (FFG), the province of Styria (SFG) and partners from industry and academia. The COMET Programme is managed by FFG.

\balance
\printbibliography

\end{document}